\pdfoutput=1
\documentclass[11pt]{article}

\usepackage{etoolbox}
\newtoggle{anon}
\togglefalse{anon}

\usepackage[final]{acl}

\usepackage{times}
\usepackage{latexsym}

\usepackage[T1]{fontenc}
\usepackage[utf8]{inputenc}

\usepackage{microtype}

\usepackage{inconsolata}

\usepackage{graphicx}

\title{NLP Meets the World: Toward Improving Conversations With News Media About Natural Language Processing Research}

\iftoggle{anon}
{
\author{Anonymized Camera-Ready Manuscript}
}
{
\author{Shomir Wilson \\
  Department of Human-Centered Computing and Social Informatics \\
  College of Information Sciences and Technology \\
  Pennsylvania State University \\
  \texttt{shomir@psu.edu}\\}
}

\begin{document}
\maketitle
\begin{abstract}

Recent advances in large language models (LLMs) have rapidly increased the public's interest in natural language processing (NLP). This attention is reflected by news outlets inviting NLP researchers to share their knowledge and views with public audiences. Recognizing the opportunities for the research field and for individual researchers, this paper shares recommendations for speaking with news media about NLP's capabilities and limitations. Drawing on the topical structure of research communication frameworks published by the US National Academies of Sciences, Engineering, and Medicine and the European Union's QUEST project, recommendations are provided for three field-appropriate communication challenges: cognitive terminology, unsustainable expectations, and ethical failures. Published NLP research and popular news coverage are cited to provide examples. The recommendations promote effective, transparent communication with the general public about NLP, in order to strengthen public understanding and encourage support for research.

\end{abstract}

\section*{Note}
\noindent \textit{Because of the subject matter, this paper cites a variety of types of sources, including news articles, unpublished essays, preprints, and whitepapers. Citations and references marked with an asterisk (i.e.,~*) represent sources that are not known to be peer-reviewed.}

\section{Introduction}
\label{sec:intro}

Public communication is a vital task for the natural language processing (NLP) research community, which faces a unique confluence of circumstances. Interest in LLMs has led news venues to publish non-technical explanations of how LLMs work~\cite{roose2023does, clarke2023_how_ai_chatbots_work}, predictions about LLMs' future abilities~\cite{waters2025_diverging_future_ai}, explorations of LLMs' social biases~\cite{rogers2025_ai_bias_multilingual}, and concerns about their impacts on employment~\cite{hayes2025_ai_jobs}. Journalists also have interviewed researchers who critique claims about artificial general intelligence (AGI)~\cite{heikkila2025_mitchell_agi_snake_oil, piquard2024_bender_parrots_lemonde}. Within scholarly literature, LLMs are sometimes claimed to be an existential threat to humanity~\cite{tait2024clipping, grozdanoff2023looming}, while some critique the plausibility or meaning of AGI~\cite{blilihamelin2025stoptreatingaginorthstar, mueller2024myth}. Regardless of stance, these activities reflect an unprecedented surge of public attention. % Regardless of stance, these topics and others reflect an unprecedented surge of public attention. %how NLP is experiencing an unprecedented surge of public attention.

The public discussion of NLP extends into culture and language, as NLP researchers contend with portrayals of AI in science fiction~\cite{osawa2022visions} and ambiguity over cognitive terms such as \textit{reason} and \textit{understand}~\cite{mitchell2023debate}. While research communication is a challenging task for many scholarly disciplines~\cite{simis2016lure}, little support is available for the acute challenges that NLP researchers currently face. This gap in support contrasts with the importance of research communication: it encourages public support for research funding, influences public policy, and engages students who may become future researchers.

This paper serves as a referenceable set of recommendations for NLP researchers interacting with news media, including interviewing with journalists and publishing commentary. It derives thematic and structural inspiration from research communication frameworks published by the US National Academies of Sciences, Engineering, and Medicine~\citeyearpar{national2017communicating} and the European Union's QUEST project~\citeyearpar{quest2020qualityindicators}. Their guidance is translated into three challenges for communicating NLP research: \textit{cognitive terminology} (\S\ref{sec:informative}), \textit{unsustainable expectations} (\S\ref{sec:learn}), and \textit{ethical failures} (\S\ref{sec:specificity}). Rather than replacing existing guidance (as reviewed in \S\ref{sec:related_work}), this paper fills a notable gap by attending to the contemporary challenges specifically relevant to NLP. %challenges specific to the NLP research community.

\section{Background and Related Work}
\label{sec:related_work}

The US National Academies created a framework for exploring the task of research communication \citeyearpar{national2017communicating}, but its guidance is intentionally broad (i.e., for all disciplines at once). This paper translates the three core challenges in their framework into challenges for NLP: \textit{the complexities of communicating science} (cognitive terminology, \S\ref{sec:informative}), \textit{communicating science in a complex, competitive communication environment} (unsustainable expectations, \S\ref{sec:learn}), and \textit{the nature of science-related public controversies} (ethical failures, \S\ref{sec:specificity}). Similarly, the European Union-funded QUEST (QUality and Effectiveness in Science and Technology communication) project identified 12 quality indicators for science communication and clustered them into three themes~\cite{quest2020qualityindicators}. Their themes are aligned with the same challenges: \textit{presentation} (cognitive terminology), \textit{connection} (unsustainable expectations), and \textit{trustworthiness} (ethical failures). No such mapping or guidance can be comprehensive, which is discussed further in Limitations, but these prior frameworks lend focus and structure to the present work.

%Because of the qualitative nature of this task, other interpretations may be possible, and we discuss this in Limitations.

% is sufficiently coarse-grained to provides guidance on research communication, covering all of the sciences, engineering, and medicine, and the need to articulate that guidance for NLP

% topics such as handling controversy, understanding one's audience, and using social media effectively. However, the 

Prior published literature on research communication provides little guidance tailored to NLP (i.e., in contrast with general guidance for researchers in any discipline). Notably, \citet{hudson2023} observe that using performance figures to describe explainable AI research alienates public audiences. A series of AAAI tutorials have focused on science communication (e.g., ~\citet{smith2025sciencecomm}), but their reach and accessibility differ from a published paper. \citet{10.1109/ICSE-SEET58685.2023.00017} designed a course-based exercise for computer science undergraduates to practice communicating software engineering research, but they focus on describing the exercise itself. Beyond computing, \citet{fontaine2019communicating} surveyed the literature on public communication by researchers within the health sciences. They noted wide use of social media and proposed a typology of strategies. These include minimizing the use of jargon, encouraging public discussion, and making information actionable, among others. \citet{kuncel2013communicating} address science communication for psychology, with a focus on making statistical methods understandable. 

% All of the above works are valuable contributions, and it is appropriate to select among them for a basic understanding of research communication. This manuscript builds upon them by focusing on the unique circumstances of NLP research, remedying a lack of field-specific guidance.

Finally, this paper follows an established practice of NLP conference papers that reflect on community challenges. Examples include \citeauthor{rogers-augenstein-2020-improve}'s recommendations for improving peer review~\citeyearpar{rogers-augenstein-2020-improve}, \citeauthor{blodgett-etal-2020-language}'s critique of how \textit{bias} is discussed~\citeyearpar{blodgett-etal-2020-language}, and \citeauthor{mosbach-etal-2024-insights}'s discussion of interpretability research~\citeyearpar{mosbach-etal-2024-insights}.

\section{Cognitive Terminology}
\label{sec:informative}

NLP researchers often describe their work using terms that public audiences associate with cognition, and the cognitive claims those terms imply might be intended or unintentional. Additionally, the research community lacks an agreed-upon stance on some cognitive claims, partly because of a lack of agreement for what they mean~\cite{bender-koller-2020-climbing}. These conditions are difficult for non-experts to navigate, limiting their abilities to interpret or support NLP research. This section illustrates with terms used in published research.

\subsection{It's Complicated: NLP's Relationship With Cognitive Terms}

Terms with cognitive implications (among their other implications) have a long history in NLP, and some are widely interpreted by researchers as non-cognitive.\footnote{Note this manuscript does not evaluate cognitive claims about LLMs. Instead, it shows how the NLP research community uses cognitive terms differently from the general public, including examples that predate LLMs.} \textit{Predict}, for example, has been associated with text classification at least as early as the 1990s~\cite{riloff1992classifying, 10.1145/183422.183425} and possibly earlier. \textit{Read} also has a long history~\cite{156474, GOVINDAN1990671}. \textit{Learn} (as in \textit{machine learning}) originates from a 1959 paper by Arthur Samuel about programs that play the game of checkers~\cite{10.1147/rd.33.0210, foote2021briefml}. However, a non-expert may associate all three of those terms with the ability to \textit{think} or \textit{understand}, broadly without definitions. Observe that ``a \textit{human} must understand to predict'' is intuitively reasonable, but several papers cited in \S\ref{sec:intro} show that ``a \textit{language model} must understand to predict'' is not universally agreed upon by the NLP research community.

LLMs' capabilities and limitations have amplified attention to cognitive terms. \textit{Hallucinate} is widely used by the research community~\cite{huang2025survey}, even if a precise meaning is not agreed upon~\cite{narayanan-venkit-etal-2024-audit}. However, the relevance of \textit{understand}, \textit{reason}, and \textit{think} are controversial~\cite{bender-koller-2020-climbing, shojaee2025illusionthinkingunderstandingstrengths}. Again, the relationships between these terms within NLP may not match the relationships intuited by an outside audience. For example, a human who is \textit{hallucinating} is presumably \textit{thinking} (albeit aberratively), but a \textit{hallucinating} LLM is not agreed upon by the research community to be \textit{thinking}. Additionally, researchers in the health sciences have criticized using \textit{hallucinate} to refer to LLM behavior as inappropriate and misleading~\cite{smith2023hallucination, ostergaard2023false}.

A further complication occurs when arguments against LLMs' cognitive abilities nominally exclude humans from having those same abilities. For example, \citet{shojaee2025illusionthinkingunderstandingstrengths} argue that language reasoning models (LRMs, closely related to LLMs) cannot \textit{think} because of ``a complete accuracy collapse'' for complex tasks. The human brain also has task complexity limits~\cite{marois2005capacity}, but in contrast with LLMs, the human brain is widely recognized as capable of \textit{thinking}.

\subsection{Recommendations}

NLP researchers speaking to the public should provide scope for cognitive terms they use. Given the above obstacles, it is helpful to tell an audience what terms such as \textit{predict} or \textit{reason} mean in a context. If a non-cognitive term can replace a cognitive one, that substitution may provide greater clarity. When benchmarks are relevant, describing them can demystify how a cognitive or quasi-cognitive claim is justified. A general goal is to avoid the possibility of unintended cognitive implications for LLMs or other human language technologies.

NLP researchers should balance correcting cognitive misconceptions about LLMs with tolerating a general audience's verbal shorthand and metaphor. Humans apply cognitive anthropomorphism to a variety of objects~\cite{boyer1996makes} such as lamps and umbrellas~\cite{edwards2022lamps}, and LLMs display far more sophisticated behavior than those objects. When cognitive terms surface in discussions about LLMs, a diplomatic strategy is to point out the verbal shorthand or metaphor, acknowledge its naturalness, and explain the actual limitations of the technology. This strategy validates a non-expert's experiences with technology while illuminating how cognitive terms can mislead.

\section{Unsustainable Expectations}
\label{sec:learn}

Predictions of the impacts of LLMs include radical improvements in biology, neuroscience, economic development, and human welfare, and the onset of global peace~\cite{amodei2024machines}. The possibility of an ``AI bubble''\footnote{There exists a lack of agreement on the distinction between contemporary AI and NLP, with some common ground and some differences. To avoid misattribution, this paper uses ``AI'' for topics where the term prevails in popular discussion.} (i.e., an unsustainable level of public interest in AI that exceeds its capabilities) is widely acknowledged~\cite{horowitch2025computersciencebubble, press2025peakai}. A bubble implies short-term benefits, such as publicity and abundant resources. However, its collapse---implied by the metaphor---creates harm through an unpredictably sharp decline in resources. This section provides examples from the history of AI and recommendations for how NLP researchers can encourage realistic expectations about the field.

\subsection{Remembering AI Winters}

Multiple AI winters are said to have occurred: Roger Schank and Marvin Minsky referred to one in the 1970s~\cite{crevier1993ai}, and James Hendler referred to one in the 1980s~\cite{4475849}. This section uses Hendler's more recent perspective to structure the discussion.

Hendler described a confluence of three conditions that led to the AI winter of the 1980s. The first was a decrease in government funding for AI research in the 1970s, specifically in the USA, which dominated AI research at the time. He argues the impact was delayed by five to ten years, as already-approved projects ran their course. For the rationale of the cuts, Hendler quotes a 1973 report by James Lighthill in the United Kingdom that concluded ``in no part of the field have discoveries made so far produced the major impact that was then promised.''~\cite{james1973artificial} The second condition was an erosion of recognition for AI's successes. Hendler argues that expert systems~\cite{buchanan1988fundamentals}, which were successful in industry, were ``disowned'' by AI researchers in favor of topics where less progress had been made. The third condition was the rapid democratization of hardware needed to run AI models, in the sense that the hardware became dramatically cheaper and easier to own, thus disrupting sociotechnical power structures within research~\cite{4475849}.

These conditions require some subjectivity to assess, making it difficult to firmly state whether they exist now or will appear in the near future. However, there are discernible lessons in them for how NLP researchers communicate their work.

\subsection{Recommendations}

For the NLP research community as a whole, the downsides of setting realistic expectations are modest when compared to the downsides of a sudden collapse in public interest. At the individual level, one researcher's careful management of expectations when speaking about their work is unlikely to change public perception of the field, but strategies exist that fulfill that goal while creating individual benefit as well. Such a strategy is described below.

Public engagement is an opportunity to support recognition of NLP applications that may have been inadvertently ``disowned'' (borrowing Hendler's description) for their success. NLP played a formative role in early search engines~\cite{schwartz1998web} and enabled early spam detection~\cite{cormack2008email}. Optical character recognition made a wealth of pre-digital texts searchable~\cite{hill2019quantifying}, and authorship identification solved historical mysteries~\cite{tweedie1996neural}. Those examples and others show that NLP can enable \textit{calm technologies}, supporting human work rather than monopolizing human attention~\cite{weiser1996designing}. They also may serve as an antidote to hype by illustrating that research products need time to mature. Dialog systems are a strong public-facing example of the long road to practical use: early systems such as Eliza~\cite{weizenbaum1966eliza} are distant ancestors of modern virtual assistants, such as Amazon Alexa, Siri, and Google Assistant.

\section{Ethical Failures}
\label{sec:specificity}
Ethically dubious practices are known to exist within the NLP research community, and external to it, problematic NLP applications also exist. Both are obstacles to how NLP research is received by a public audience. This section gives examples of relevant ethical failures and provides suggestions for how NLP researchers can speak frankly and constructively when called upon to discuss them.

\subsection{Reckoning With Problematic NLP}

Research institutions in many countries have procedures for reviewing experimental protocols that involve human subjects. However, sometimes an experiment complies with requirements while violating ethical norms, or researchers fail to seek approval. A recent example is an experiment in which researchers studied the persuasiveness of an LLM-powered chatbot by covertly deploying it on the Reddit forum \textit{r/changemyview}\footnote{https://www.reddit.com/r/changemyview/}, concealing that it was an LLM. When forum moderators were informed about the deception, they expressed shock, asked for an apology from the university, and requested that the research not be published~\cite{OGrady2025_unethical_AI_Reddit}. The experiment also reflects a mismatch between researchers' expectations and a studied population's expectations for ethical uses of NLP. Another example of mismatched ethical expectations is the use of de-identified conversation data from Crisis Text Line, a free nonprofit service for people in crisis in the US, by a for-profit company. Although users of Crisis Text Line were provided with a link to terms of service, informing them how their data could be shared, public backlash led to the termination of the data sharing agreement~\cite{eysenbach2025crisis}.

A comprehensive list of ethical failures or misuses of NLP is impractical, but examples are readily available. LLMs have been used to generate disinformation, potentially manipulating geopolitics~\cite{doi:10.1177/0163443716686672}. Chatbots have demonstrated the capacity to respond harmfully in conversations with people suffering from mental illness~\cite{coghlan2023chat}, and more broadly, LLMs' sociodemographic biases have been shown to harm individuals from minoritized groups~\cite{Ghosh2024}. The broad-based collection of training data for LLMs raises privacy concerns and intellectual property concerns~\cite{novelli2024generative}. NLP and computing security are increasingly intertwined, obligating NLP researchers to learn more about the nuanced ethics of security research~\cite{lent-etal-2025-nlp}. LLM-assisted writing has been claimed to produce ``cognitive debt'' and reduce critical thinking~\cite{kosmyna2025brainchatgptaccumulationcognitive}. LLM training and usage consume significant resources, raising questions about environmental sustainability~\cite{kogkalidis-chatzikyriakidis-2025-tables}. Finally, NLP research is vulnerable to distortion by monied interests, potentially influencing individual researchers and entire institutions~\cite{10.1145/3461702.3462563}.

The above harms and others receive public attention, to the extent that a 2023 Pew Research survey found 52\% of Americans were more concerned than excited about AI in their daily lives~\cite{FaverioTyson2023}. This level of concern is demonstrably relevant to NLP researchers.

\subsection{Recommendations}

Several popular news articles cited in this paper show that NLP researchers are called upon to discuss problematic uses of NLP. Open dialog about these problems is obligatory and familiar as part of open science, but further steps can help to sustain public support. Researchers speaking with the general public about problematic NLP should be prepared to explain what should have been done differently to achieve the same goals, or why the goals never should have been undertaken. Public interaction is also an opportunity to point to research directly aimed at solving ethical issues in NLP, showing how human-centered research is a valuable part of the community. Such research is a marked contrast with a hypothetical public image of AI research as solely focused on computation.

Within the research community, researchers can lay the groundwork for strong public support by recognizing that research that critiques NLP fully belongs in the community. Such research does not necessarily create human language technologies, but it guides their ethical development. This inclusion also encourages grounding NLP research in human benefit, which demonstrates responsiveness to the concerns of the general public.

\section{Conclusion}

Beyond this paper, other strategies are possible for encouraging NLP researchers to engage productively with news media. Among them are holding media training sessions at conferences and developing curricular modules for students to learn about research communication. A multi-pronged strategy, including these activities or others, has the greatest potential to reach a wide audience of researchers.

Public attention to NLP research comes with benefits, obstacles, and responsibilities. This paper presented recommendations to help NLP researchers communicate effectively with the public, focusing on obstacles that are distinctive to our research community. It drew upon a variety of sources to illustrate those obstacles and how to navigate them, for the prosperity of individual researchers and for the research community at large.

\iftoggle{anon}{}{

\section*{Acknowledgments}

The reflections in this manuscript benefited from several experiences discussing research with news media, and I gratefully acknowledge the work of Penn State's Office of Strategic Communications and (also within Penn State) the College of IST's Office of Marketing and Communications, who made those media engagements possible. I also gratefully acknowledge Kenneth Huang and Natalie Parde for their feedback on ideas in this manuscript. This manuscript is based upon work supported by the National Science Foundation under Grant \#2237574.

} % close anon iftoggle

\section*{Limitations}

A comprehensive guide to research communication, examining all possible scenarios, is infeasible. Instead, this manuscript augments existing discipline-nonspecific guides to focus on NLP. In \S1 and \S2 readers are reminded of that, and in \S2 more sources are cited for the reader to consult. Additionally, the complexity and subjectivity of research communication mean that no one guide (or even the union of all guides) will be comprehensive for the problems researchers face. However, other fields appear to find research communication guides helpful, as shown by their acceptance for publication, the citations to them, and reflections on them~\cite{doi:10.1073/pnas.1212729110}.

NLP researchers may face public communication challenges not covered by this paper's guidance. This paper identifies challenges of particular concern by interpreting established research communication frameworks, but other challenges may exist. The subjective nature of the topic space obviates covering it in an exhaustive way, but sharing guidance grounded in recognized prior frameworks (as this manuscript does) is deemed more likely to lead to positive outcomes than the relative lack of guidance in the status quo.

\section*{Ethical Considerations}

This paper is reflective in purpose and does not represent experiments, and its ethical ramifications are limited to the guidance it provides. Existing research communication frameworks and other cited literature provide evidence that the topics in this manuscript are relevant and have received little or no prior attention for NLP. The reader is encouraged to consult additional guidance as necessary.

%Since December 2023, a "Limitations" section has been required for all papers submitted to ACL Rolling Review (ARR). This section should be placed at the end of the paper, before the references. The "Limitations" section (along with, optionally, a section for ethical considerations) may be up to one page and will not count toward the final page limit. Note that these files may be used by venues that do not rely on ARR so it is recommended to verify the requirement of a "Limitations" section and other criteria with the venue in question.

\bibliography{aclrr2025_shomir}

\end{document}